\documentclass[
 reprint,
 amsmath,amssymb,
 aps, floatfix, prl
]{revtex4-2}

\usepackage{graphicx}
\usepackage{dcolumn}
\usepackage{bm}
\usepackage{aas_macros}
\usepackage{hyperref}
\usepackage[dvipsnames]{xcolor}
\usepackage{ulem}

\begin{document}

\preprint{XXX}

\title{Bounding the mass of ultralight bosonic Dark Matter particles\\with the motion of the S2 star around Sgr A*}
\author{Riccardo Della Monica}
\email{rdellamonica@usal.es}
\affiliation{%
    Departamento de F\'isica Fundamental, Universidad de Salamanca,\\Plaza de la Merced, s/n, E-37008 Salamanca, Spain
}%
\author{Ivan de Martino}%
\email{ivan.demartino@usal.es}
\affiliation{%
    Departamento de F\'isica Fundamental and IUFFyM, Universidad de Salamanca,\\Plaza de la Merced, s/n, E-37008 Salamanca, Spain
}%

\date{\today}% 

\begin{abstract} 
    Dark matter is undoubtedly one of the fundamental, albeit unknown, components of the standard cosmological model. The failure to detect WIMPs, the most promising candidate particle for cold dark matter, actually opens the way for the exploration of viable alternatives, of which ultralight bosonic particles with masses $\sim 10^{-21}$ eV represent one of the most encouraging. Numerical simulations have shown that such particles form solitonic cores in the innermost parts of virialized galactic halos that are supported by internal quantum pressure on characteristic $\sim$kpc de Broglie scales. In the Galaxy, this halo region can be probed by means of S-stars orbiting the supermassive black hole Sagittarius A* to unveil the presence of such a solitonic core and, ultimately, to bound the boson mass $m_\psi$. Employing a Monte Carlo Markov Chain algorithm, we compare the predicted orbital motion of S2 with publicly available data and set an upper bound $m_\psi \lesssim 3.2\times 10^{-19}$ eV on the boson mass, at 95 \% confidence level. When combined with other galactic and cosmological probes, our constraints help to reduce the allowed range of the bosonic mass to $(2.0 \lesssim m_\psi \lesssim 32.2)\times 10^{-20}$ eV, at the 95 \% confidence level, which opens the way to precision measurements of the mass of the ultralight bosonic dark matter.
\end{abstract}

\maketitle

\textit{Introduction} --- The currently accepted cosmological model, built upon the theory of General Relativity (GR), prescribes that $\sim$27\% of the matter-energy content of the Universe resides in the form of a 
yet unknown Dark Matter (DM) component \cite{deMartino2020}. Hierarchical structure formation \cite{Springel2005} and the spectrum of CMB fluctuations \cite{Planck2020} point towards a Cold DM (CDM) scenario, according to which DM is composed of non-relativistic particles 
belonging to physical scenarios beyond the Standard Model of particle physics, and weakly interacting with ordinary matter solely through gravitational forces \cite{Salucci2021}. Despite the success of CDM-based cosmological simulations in predicting large scale structures in the Universe, the CDM model is currently challenged by a remarkable lack of detection of
its most promising candidate, {\textit i.e.} the WIMPs, in Earth-based laboratories \cite{Aprile2018, CMS2019} and by seemingly unreconcilable problems on galactic or sub-galactic scales (\textit{i.e.} $< 10$ kpc) \cite{Moore1994, Klypin1999, deBlok2010}. Contrary to what observations suggest, especially in dwarf 
and low surface brightness galaxies, the lack of internal pressure in self-gravitating CDM halos would result in inherently spiky density profiles, and to the absence of a natural mechanism that prevents the formation of structures below a certain scale \citep{deMartino2020}. This gives rise to the well-known ``missing satellite'', ``too big to fail'' and ``core-cusp'' problems \cite{Moore1994, Klypin1999, deBlok2010}. Some studies invoke tidal forces and a not completely understood baryonic feedback mechanism (from supernovae and active galactic nuclei) to alleviate the tension between simulations and observations \cite{Brooks2013, DelPopolo2016}. Nonetheless, 
since the aforementioned proposals have not been validated and established yet, other paths have been investigated 

Wave Dark Matter 
(commonly referred to also as Fuzzy Dark Matter or $\psi$-Dark Matter) is considered among the most promising alternatives \cite{Hui2017} as in it all such seemingly contradicting 
astrophysical observations find a natural explanation \cite{Marsh2015}. 
In this scenario, DM is composed of ultralight particles ($m_\psi < 10^{-17}$ eV), whose macroscopic de Broglie wavelength $\lambda_{\rm dB} \sim 1$ kpc gives rise to quantum behaviors on astrophysical scales and whose existence, at a fundamental level, seems to be justified by axionic DM particles generated through a symmetry breaking by the misalignment mechanism \cite{Abbott1983, Preskill1983} 
rather common in string landscape \cite{Arvanitaki2010}.
Inherently spinless, such particles follow a Bose-Einstein statistics \cite{Sikivie2009}.
As a consequence of that, a self-gravitating halo composed of such particles builds an internal quantum pressure, sustained by the uncertainty principle. This prevents wave DM to be confined on scales smaller than its de Broglie wavelength, thereby providing a natural lower 
Jeans scale for structure formation and leading to the emergence of a prominent soliton core \cite{Davies2020}, as confirmed by pioneering cosmological simulations in \cite{Schive2014}. The generation of coherent standing waves of wave DM in the outskirts of galactic halos leads to the formation of density fluctuations, owing to quantum interference patterns on the de Broglie scale \cite{Schive2014b}, that, once azimuthally averaged, follow a Navarro-Frenk-White (NFW) profile \cite{Schive2014,Mocz2017}. Nonetheless, on much larger cosmological scales wave DM behaves exactly as CDM would, thus predicting the same large scale structures and overall cosmic evolution \cite{Hui2017}. 
Several efforts have been made to design astrophysical tests that could directly constrain the mass of the boson $m_\psi$. The most recent one claimed the direct observations of the effects of the quantum interference pattern in the anomalies of gravitational lensing images \cite{Amruth2023}. Previously, Jeans analysis of local dwarf spheroidal galaxies \cite{Chen2017, Pozo2020, Pozo2023} revealed the natural ability of wave DM to reproduce cored density profiles and the reported transition in stellar density at $\sim 1$ kpc 
from the center of the galaxies. Moreover, stellar velocity dispersion profiles consistent with the presence of a wave DM solitonic core, are found to reproduce the behavior of low-density self-gravitating systems \cite{Broadhurst2020} such as ultra-diffuse galaxies \cite{Pozo2021}.  
Additionally, the velocity dispersion of bulge stars 
within the first hundreds of pc from the center of the Milky Way (MW) has been shown to be consistent with the presence of 
a solitonic DM core with a typical scale length of $\sim 300$ pc,
favouring a boson mass $10^{-22}$ eV \cite{deMartino2020b}. 

Deeper towards the center of the MW, in \cite{DellaMonica2022c} the effects of the presence of a wave DM soliton were considered on the motion of the S-stars around the four-million solar masses supermassive black hole (SMBH) Sagittarius A* (Sgr A*) in the Galactic Center (GC) of the MW \cite{deLaurentis2022} (previously, such stars also helped providing constrains on NFW-like spiky dark matter distributions in the GC \cite{Lacroix2018, Shen2023}). The sensitivity analysis in \cite{DellaMonica2022c}, performed upon applying a post-Newtonian (PN) approximation, showed that the astrometric observations of the orbit of the brightest star in the cluster, S2, are sensible to boson masses below $10^{-19}$ eV, thus providing a possible new avenue to test the wave DM model and to narrow the range of allowed values for $m_\psi$.

\begin{figure*}
    \centering
    \includegraphics[width = \textwidth]{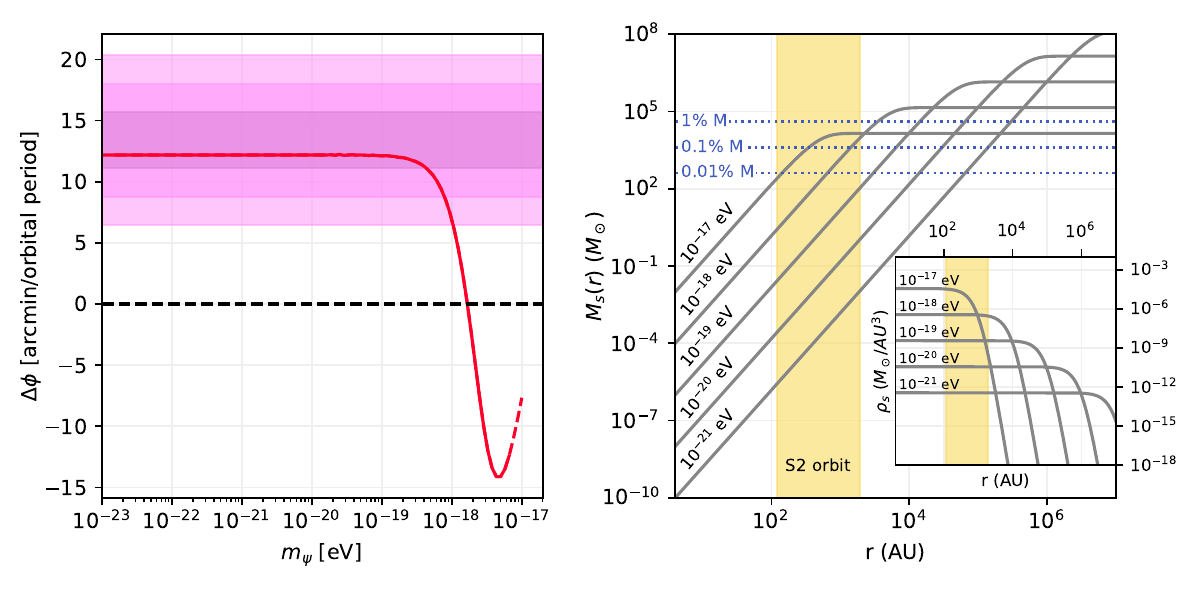}
    \caption{\textit{Left panel:} Dependence of the orbital precession for the S2 star from the boson mass $m_\psi$ in the range of interest $m_\psi\in[10^{-23},10^{-17}]$ eV. The pink horizontal bands report the $1\sigma$, $2\sigma$ and $3\sigma$ experimental bounds imposed by the Gravity Collaboration from analysis of the orbital data for S2 \cite{Gravity2020}. \textit{Right panel:} Enclosed mass $M_s(r)$ of the solitonic wave DM distribution as a function of radius in units of solar masses, for different values of the boson mass $m_\psi$ as reported in the corresponding labels. The yellow shaded strip higlights the extent of the S2 star orbit. Blue dotted lines report values of the enclosed mass corresponding to 1\%, 0.1\% and 0.01\% of Sgr A* mass $M$, respectively. The inset plot shows the solitonic density profiles in Eq. \eqref{eq:soliton} for the same different values of the boson mass $m_\psi$.}
    \label{fig:horizons_precession}
\end{figure*}

\textit{The model} --- 
With the aim of providing the first constraint of wave DM within the first pc from the center of the MW, here we study the geodesic motion of the S2 star
by considering a space-time metric that accounts for the simultaneous presence of the SMBH and of the wave DM soliton. The solitonic condensate 
follows the empirical profile \cite{Schive2014}
\begin{equation}
    \rho_s(r) = \rho_c\left(1+\alpha\left(\frac{r}{r_c}\right)^2\right)^{-8},
    \label{eq:soliton}
\end{equation}
where $\alpha=\sqrt[8]{2}-1\approx 0.091$ and the central density $\rho_c$ depends on the boson mass $m_\psi$ via
\begin{equation}
    \rho_c = 0.019 \left(\frac{m_\psi}{10^{-22}\textrm{eV}}\right)^{-2}\left(\frac{r_c}{1\textrm{ kpc}}\right)^{-4}\frac{M_\odot}{\rm pc^3},
\end{equation}
with the solitonic core radius $r_c$ satisfying the scaling relation \cite{Schive2014b}
\begin{equation}
    r_c = 1.6\left(\frac{m_\psi}{10^{-22}\textrm{ eV}}\right)^{-1}\left(\frac{M_{\rm halo}}{10^9 M_\odot}\right)^{-1/3}\textrm{kpc},
    \label{eq:scaling_relation}
\end{equation}
being $M_{\rm halo}$ the mass of the entire DM halo. Since for the MW the halo mass is known to be $M_{\rm halo} = 1.08 \times 10^{12} M_\odot$ by GAIA satellite observations \cite{Cautun2020}, the solitonic profile in Eq. \eqref{eq:soliton} depends solely on the boson mass $m_\psi$.
The core-halo relation in Eq. \eqref{eq:scaling_relation} has been thoroughly investigated in \cite{Chan2022}, where dependence on the simulation setup has been reported. In particular, a sizeable dispersion in the core–halo mass relation is found for higher halo masses, pointing towards heavier solitons. Since a heavier soliton would alter more drastically the stellar dynamics, we take a conservative approach and assume Eq. \eqref{eq:scaling_relation} to be valid in our case.
In \cite{Pantig2023}, an analytic space-time solution is constructed from the solitonic mass profile in Eq. \eqref{eq:soliton} assuming that the wave DM particles surround the BH in a spherically symmetric configuration, that the presence of the SMBH does not alter significantly the wave DM profile, {and that we can neglect accretion of the DM soliton on the central SMBH. Regarding the latter two assumpions, in \cite{Davies2020} it has been shown that for $m_\psi \gtrsim 10^{-20}$ eV the solitonic profile in Eq. \eqref{eq:soliton} might indeed be affected by the presence of the SMBH, leading to a more compact and massive soliton. This, on turn, would affect more drastically the S2 dynamics. For this reason, we take a conservative approach and consider the solitonic profile unaltered by the presence of Sgr A*.
Following \cite{Xu2018} (whose approach led to the derivation of other metric solutions for several known DM profiles and alternative BHs \cite{Daghigh2022,Liu2021,Jusufi2020}), the wave DM density profile in Eq. \eqref{eq:soliton} is introduced in the energy-momentum tensors in the Einstein's field equation, leading to a 
stationary and spherically symmetric BH solutions embedded in a wave DM halo 
\cite{Pantig2023}
\begin{equation}
    ds^2 = -F(r)dt^2+\frac{1}{F(r)}dr^2+r^2(d\theta^2+\sin^2\theta d\phi^2),
    \label{eq:metric}
\end{equation}
where
\begin{equation}
    F(r) = \exp\left(\frac{4\pi \rho_c r_c^3}{91 \alpha^{8} r}\left(\left(\frac{r_c}{r}\right)^{13}\mathcal{F}-\frac{3003 \pi\alpha^{\frac{13}{2}}}{2048}\right)\right)-\frac{2M}{r},
    \label{eq:metric_function}
\end{equation}
being $\mathcal{F}$ an hypergeometric function of the radial coordinate $r$, given by
\begin{equation}
    \mathcal{F}(r) = {}_2F_1\left(\frac{13}{2},\,7;\frac{15}{2};\,-\frac{r_c^2}{r^2\alpha} \right).
    \label{eq:hypergeometric}
\end{equation} 
The positive root of the function in Eq. \eqref{eq:metric_function} identifies the radial coordinate $r_H$ of the event horizon. We derive $r_H$ numerically as a function of the wave DM boson mass $m_\psi$ in the range of interest of ultralight axions $m_\psi \in [10^{-23},10^{-17}]\textrm{ eV}$, and find a negligible variation ($<10^{-6}M$) with respect to the Schwarzschild case, $r_H^{\rm Sch} =  2GM/c^2$. Deviations of the horizon radius, leading to a comparatively larger size of the BH shadow for the metric in Eq. \eqref{eq:metric_function}, have been examined in \cite{Pantig2023}, as a function of both the boson mass and of the core radius $r_c$, assumed as independent parameters (\textit{i.e.} without taking into consideration the scaling relation in Eq. \eqref{eq:scaling_relation}).
Due to the current low-resolution of event horizon imaging \cite{EHT2022a}, however, such horizon-scale tests alone are currently unable to place sensible bound on the wave DM particle mass when the scaling relation in Eq. \eqref{eq:scaling_relation} is taken into account.
Nonetheless, greater effects are expected on bigger scales, due to the larger fraction of enclosed integrated mass of the wave DM distribution. Stronger effects of weak lensing and a larger Einstein ring due to the presence of the soliton effects are obtained in \cite{Pantig2023}, especially when the impact parameters is comparable to the soliton core. 

\textit{Methodology} --- Following \cite{DellaMonica2022a,DellaMonica2022b,DellaMonica2023a}, we integrate 
numerically the time-like geodesic equations derived from the space-time metric in Eq. \eqref{eq:metric}, to predict theoretical orbits for the S2 star around Sgr A*. Initial conditions for the orbit are assigned in terms of the classical Keplerian elements: the semi-major axis $a$, the eccentricity $e$, the time of pericenter passage $t_p$ and the orbital period $T$ uniquely identify a Keplerian ellipse on the equatorial plane (which due to the spherical symmetry of our configuration does not lead to a loss of generality), that we assume to osculate the real trajectory of the star at the initial time (which we set to the time in which S2 last passed at its apocenter $t_0\sim$2010.35). 
We hence employ a fourth-order Runge-Kutta scheme, to integrate the geodesic equations over approximately two orbital periods (corresponding to a time range spanning from $\sim$1990 to $\sim$2020). This allows us to compute numerically the relativistic periastron advance experienced by the star and compare it with the one induced by a Schwarzschild BH of mass $M$ whose 1PN expression is \cite{Poisson2014}
\begin{equation}
    \Delta\phi = \frac{6 \pi G M}{a c^2 (1-e^2)}
    \label{eq:precession}
\end{equation}
for each orbital period. Considering a mass $M \simeq 4.2\times 10^6 M_\odot$ for Sgr A* and the orbital parameters found in literature \cite{Gravity2020} for S2, this angle amounts to $\sim 12.1'$. The presence of an extended mass component around Sgr A* is expected to induce an additional periastron shift referred to as \textit{mass precession} \cite{Heissel2022} that is demonstrated to counteract the prograde Schwarzschild precession. In the left panel of Fig. \ref{fig:horizons_precession}, we report the results of our numerically estimated rate of orbital precession for S2 form the geodesic integration as a function of the wave DM boson mass $m_\psi$. The pink bands in the figure report the 1$\sigma$, 2$\sigma$ and 3$\sigma$ bounds on the measured rate of orbital precession for S2 obtained by the precise monitoring of its orbital motion by the Gravity Collaboration \cite{Gravity2020}. The presence of the wave DM solitonic core becomes more prominent at around $m_\psi\sim 10^{-19}$, when the orbital precession starts to decrease noticeably form the value of 12.1'/orbital period expected in a purely BH metric. At $m_\psi \sim 10^{-18}$ eV the deviation is so strong that the orbital precession would exceed the $3\sigma$ bounds imposed by astrometric observations, and at even greater masses the wave DM mass precession effect becomes dominant over the prograde Schwarzschild precession, with the overall effect becoming retrograde from $m_\psi\sim 1.5\times10^{-18}$ eV on. This prediction agrees with the previous result from a PN approximation reported in \cite{DellaMonica2022c} (where a similar dependence of the astronomical observables from the boson mass is reported) and with the constraints on the possible dark mass enclosed in the S2 orbit at around 0.1\% of the Sgr A* mass \cite{Heissel2022,Gravity2019}. 
This is shown in the right panel of Fig. \ref{fig:horizons_precession}, in which we report both the enclosed mass $M(r)$ and the solitonic density profile $\rho(r)$ (inset plot) for different values of the boson mass $m_\psi$. The region corresponding to the S2 orbit is highlighted by a shaded yellow strip. Evidently, for boson masses greater than $\sim10^{-19}\textrm{ eV}$ the mass of the solitonic distribution enclosed within the S2 orbit reaches values ranging form 0.1\% to 1\% of the mass of Sgr A*. At the same time, for the same boson masses, the density profiles within the S2 orbit switch from a flat distribution to a steeply ($\rho\propto r^{-16}$ 
) decaying one within the extent to the orbit itself. The combination of significant enclosed mass and steep density gradient within the orbit are responsible for a modification of orbital dynamics \cite{Heissel2022} that is consistent with our periastron advance results in the considered range of masses.
As such, the results reported here, and the intrinsic non-linearity of the $\Delta\phi - m_\psi$ dependence, demonstrate the importance of the orbital precession of the S2 star as a quantitative probe on the presence of a wave DM solitonic core in the GC and on its corresponding boson mass $m_\psi$, that we aim to leverage to statistically bound $m_\psi$.

An additional possible relativistic observable is the gravitational redshift, produced by the central source, on the light emitted by the star. This effect (together with the special relativistic transverse Doppler) accounts for an additional $\sim 200$ km/s in the spectroscopically reconstructed line-of-sight velocity for the S2 star at its pericenter \cite{gravity2018,Do2019}. From our numerically integrated orbits for S2, we have estimated that this effect shows very little dependence on the boson mass (with deviations below 0.2\% with respect to the GR value in the considered range of masses). This is in accordance with the fact that the wave DM induced modifications to the BH metric do not alter significantly the compactness of the central object, owing to the negligible variation in the event horizon radius.

\textit{Data and data analysis} --- In order to impose constraints on the boson mass $m_\psi$ we fit the predicted orbits for the S2 star 
to the publicly available astronomical data. 
Such a data set consists of astrometric sky-projected positions (right ascension, $\alpha$, and declination, $\delta$) 
\cite{Gillessen2017} recorded in the near-infrared at $N_p = 145$ epochs over the course of the last three decades (from $\sim$1992 to $\sim$2017), referred to the ``GC radio-to-infrared reference system'' \cite{Plewa2015} and of $ N_v= 44$ spectroscopically derived measurements of the line-of-sight velocity ($v_{\rm LOS}$) in \cite{Gillessen2017} covering approximately the same temporal period. Crucially, these data miss the last S2 pericenter passage which took place in May 2018 and the subsequent motion, affected by the periastron advance, which was indeed observed by the Gravity Collaboration \cite{Gravity2020, Gravity2021}. However, information on the measured rate of precession is encoded in a parameter $f_{\rm SP}$ (where $f_{\rm SP} = 0$ corresponds to a non-preceeding ellipse consistent with Newtonian gravity and $f_{\rm SP} = 1$ corresponds to the purely-BH general-relativistic rate of orbital precession in Eq. \eqref{eq:precession}), which was bound at $1\sigma$ to be $f_{\rm SP}= 1.10\pm0.19$, thus excluding Newtonian motion at $5\sigma$ \cite{Gravity2020}. Following the prescription from previous works \cite{deMartino2021,DellaMonica2022a}, we thereby considered the measurement of $f_{\rm SP}$ as an additional data point alongside the astrometric and spectroscopic observations. In order to reconstruct these observable quantities from the numerically computed orbits, several additional parameters are introduced: the distance $D$ of Earth from the GC; three angular orbital elements, namely the orbital inclination $i$ the longitude of the ascending node $\Omega$ and the argument of the pericenter $\omega$ required to project the orbit from the BH equatorial plane to the celestial sphere of an Earth-based observer; and five zero-point offset and drift parameter for the astrometric reference frame $x_0$, $y_0$, $v_{x,0}$, $v_{y,0}$ and $v_{z,0}$ (we refer to previous works \cite{deMartino2021, DellaMonica2022a,DellaMonica2022b} for a more thorough description of all such parameters and how they are implemented in our orbital model). All these parameters make up the 14-dimensional parameter space that we have explored by means of a Markov Chain Monte Carlo (MCMC) algorithm implemented in \cite{ForemanMackey2013}. Uniform priors are assigned to all the Keplerian parameters of our orbital model (in intervals centered on the best-fit values from \cite{Gillessen2017}, with amplitudes being 10 times those of the corresponding credible intervals) and to the parameter $m_\psi$ in the range of interest $m_\psi\in[10^{-23},10^{-17}]$ eV that we have sampled logarithmically. Conversely, for the reference frame parameters Gaussian priors have been assigned, resulting from the independent analysis in \cite{Plewa2015}. We have adopted the following likelihood
\begin{align}
	\log \mathcal{L} =& -\frac{1}{2}\biggl[\sum_i^{N_p}\biggl(\frac{\alpha_i-\alpha^{{obs}}_{i}}{\sqrt{2}\sigma_{\alpha,i}}\biggr)^2+\sum_i^{N_p}\biggl(\frac{\delta_i-\delta^{{obs}}_{i}}{\sqrt{2}\sigma_{\delta,i}}\biggr)^2+\nonumber\\
	& \sum_i^{N_v}\biggl(\frac{v_{{\rm LOS}, i}-v_{{\rm LOS}, i}^{obs}}{\sqrt{2}\sigma_{v_{\rm LOS},i}}\biggr)^2+\biggl(\frac{f_{\rm SP}-f_{{\rm SP}}^{{obs}}}{\sqrt{2}\sigma_{f_{\rm SP}}}\biggr)^2\biggr],
	\label{eq:likelihood}
\end{align}
where superscripts $obs$ stand for the observed quantities at the $i\text{-th}$ epoch, while quantities without a superscript refer to our predicted orbits. The $\sigma$s are the corresponding observational uncertainties. As first done in \cite{deMartino2021}, we introduce a factor $\sqrt{2}$ in the denominators to avoid double-counting data points when considering the last term with the orbital precession (that has partially been derived from the same dataset).

\begin{table}
    \setlength{\tabcolsep}{20pt}
    \renewcommand{\arraystretch}{1.5}
    \begin{tabular}{lr}
        \hline
        Parameter & Value \\ \hline 
        $D$ (kpc) & $8.03\pm0.21$ \\
        $T$ (yr) & $16.026\pm0.027$ \\
        $t_p-2018.38$ (yr) & $-0.016\pm0.023$ \\
        $a$ (as) & $0.126\pm0.0011$ \\
        $e$ & $0.8842_{-0.0024}^{+0.0023}$ \\
        $i$ ($^\circ$) & $133.8\pm0.48$ \\
        $\Omega$ ($^\circ$) & $226.75_{-0.78}^{+0.79}$ \\
        $\omega$ ($^\circ$) & $65.53_{-0.74}^{+0.73}$ \\
        $x_0$ (mas) & $0.33\pm0.15$ \\
        $y_0$ (mas) & $-0.01\pm0.19$ \\
        $v_{x,0}$ (mas/yr) & $0.067_{-0.051}^{+0.052}$ \\
        $v_{z,0}$ (mas/yr) & $0.12_{-0.064}^{+0.063}$ \\
        $v_{z,0}$ (km/s) & $-5.8\pm4.5$ \\
        $m_\psi$ (eV) & $\lesssim 3.2\times 10^{-19}$ {\footnotesize (95\% c.l.)} \\ \hline
    \end{tabular}
    \caption{The results of our posterior analysis for the 14 parameters of our orbital model. All the orbital parameters are bounded (the reported limits comprise a 68\% confidence interval around the median) while for the boson mass $m_\psi$ (last row) we derive an upper limit at 95\% confidence level.}
    \label{tab:posterior}
\end{table}

\begin{figure}
    \centering
    \includegraphics[width = \columnwidth]{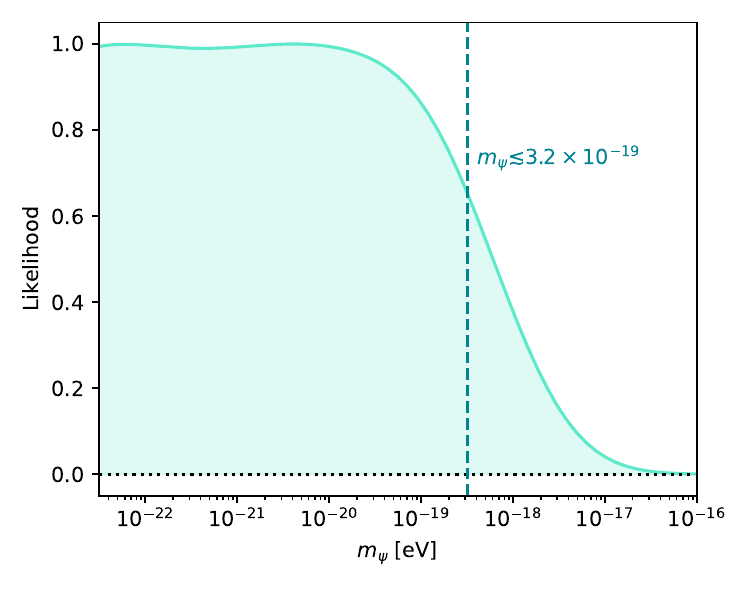}
    \caption{Posterior distribution for the boson mass $m_\psi$ from our MCMC analysis. The vertical dashed line, corresponding to the value reported in the label of $m_\psi\sim3.2\times10^{-19}$ eV, represent the 95\% confidence level upper limit for the boson mass from our analysis.}
    \label{fig:posterior}
\end{figure}

\textit{Results and Discussions} --- We present the results of our MCMC analysis in Table \ref{tab:posterior}. Median values and 68\% confidence level intervals around it are reported for the parameters that result bounded from our analysis. In particular, for all the Keplerian orbital parameters and for the reference frame it is possible to derive a confidence interval that is compatible within 1$\sigma$ with previous results on the same dataset \cite{Gillessen2017}. On the other hand, we are able to place an upper limit on the boson mass $m_\psi \lesssim 3.22\times 10^{-19}$ eV at 95\% confidence level, which is consistent with both the qualitative bounds derived from the results for the orbital precession reported in the left panel of Fig. \ref{fig:horizons_precession} and with the putative bounds imposed on $m_\psi$
using a PN fit to mock data in \cite{DellaMonica2022c}. For this parameter, we additionally report the marginalized likelihood in Fig. \ref{fig:posterior}.  

\textit{Conclusions}  --- The physical nature of DM is known to reside beyond the Standard Model of particle physics. In this context, ultralight axions \cite{Arvanitaki2010} represent a well-motivated DM candidate (here dubbed wave DM) due to their natural ability to solve inherent small-scale ($<$ 10 kpc) problems of the usual CDM scenario while reproducing the same large scale cosmic evolution \cite{Marsh2015, Hui2017}. The rich non-linear dynamics of wave DM self-gravitating halos \cite{Schive2014}, regulated by quantum mechanics, brings distinctive and testable predictions that have fueled interest in this model. Several tests have been conducted to derive observational bounds on the boson mass $m_\psi$ (whose value regulates completely the evolution of wave DM structures). Cosmological analyses, for example, constrain the value of $m_\psi$ in a quite large range of $10^{-24} \lesssim m_\psi \lesssim 10^{-17}$ \cite{Hlozek2015, Hlozek2018}, while analyses on the superradiance instability \cite{Davoudiasl2019} identify the intervals $m_\psi\lesssim2.9\times 10^{-22}$ eV and $m_\psi\gtrsim4.6\times 10^{-22}$ eV as allowed regions for wave DM particles.
Conversely, other astrophysical tests have allowed placing lower limits on the boson mass, the most stringent of which, $m_\psi \gtrsim 2\times10^{-20}$ eV, comes from the analysis of the Lyman-$\alpha$ forest \cite{Rogers2021}. In this letter, we have developed a new astrophysical test to bound the wave DM particle mass $m_\psi$ within the first pc from the center of the MW, based on the orbit of the S2 star around the SMBH Sgr A*. The space-time metric of a SMBH embedded in a wave DM solitonic core, Eq. \eqref{eq:metric}, allowed us to describe the motion of S2 with a geodesic formalism and to derive its relativistic periastron advance as a function of $m_\psi$ (Fig. \ref{fig:horizons_precession}). We have adopted a conservative approach and assumed that the soliton profile in Eq. \eqref{eq:soliton} is unaltered by the presence of the SMBH, that the core-halo relation in Eq. \eqref{eq:scaling_relation} holds for the entirety of the parameter space considered and that accretion on the central SMBH of the wave DM halo is negligible. Such assumptions allowed us to reduce the free parameters only to the boson mass $m_\psi$ but at the same time they represent the main limitations of the present work that could be overcome in a future study combining our approach with the results from simulations of wave DM halos in the presence of a central SMBH. Finally, we have fitted our orbital model for S2 to publicly available data by means of an MCMC algorithm. Quite remarkably, our analysis resulted in an upper limit $m_\psi\lesssim 3.22 \times 10^{-19}$ eV for the boson mass, that, combined with the other aforementioned astrophysical tests, constrains the values of $m_\psi$ to an unprecedented level, by narrowing the range of allowed mass to only one order of magnitude. In particular, combining our result with the 95\% constraint $m_\psi \gtrsim 2\times10^{-20}$ eV from the analysis of the Lyman-$\alpha$ forest \cite{Rogers2021}, results in an allowed mass range at 95\% confidence level of $(2.0 \lesssim m_\psi \lesssim 32.2)\times 10^{-20}$ eV.
\\

\textit{Acknowledgements}  --- RDM acknowledges support from Consejeria de Educación de la Junta de Castilla y León. IDM acknowledges support from Grant IJCI2018-036198-I  funded by MCIN/AEI/10.13039/501100011033 and, as appropriate, by “ESF Investing in your future” or by “European Union NextGenerationEU/PRTR”. IDM and RDM also acknowledge support from the  grant PID2021-122938NB-I00 funded by MCIN/AEI/10.13039/501100011033 and by “ERDF A way of making Europe”. 

\bibliography{biblio}
\end{document}